\documentclass{article}

\usepackage{arxivsimplified}

\usepackage[utf8]{inputenc} 
\usepackage[T1]{fontenc}    
\usepackage{hyperref}       
\usepackage{url}            
\usepackage{booktabs}       
\usepackage{amsfonts}       
\usepackage{nicefrac}       
\usepackage{microtype}      
\usepackage{lipsum}

\usepackage{graphicx}
\graphicspath{{figures_pdf/}}
\usepackage{subfigure}
\usepackage{cite}
\usepackage[dvipsnames]{xcolor}
\usepackage{amsmath}
\usepackage{blindtext}
\usepackage{amsfonts}
\usepackage{multirow}
\usepackage{array}
\usepackage{mathtools}
\usepackage{listings}
\usepackage{xcolor}
\definecolor{codegreen}{rgb}{0,0.6,0}
\definecolor{codegray}{rgb}{0.5,0.5,0.5}
\definecolor{codepurple}{rgb}{0.58,0,0.82}
\definecolor{backcolour}{rgb}{0.95,0.95,0.92}

\usepackage{color}
\definecolor{rev1}{rgb}{0,0,0}
\definecolor{blue}{rgb}{0,0,0}

\usepackage{esvect}
\usepackage{tabularx}
\newcolumntype{L}[1]{>{\raggedright\arraybackslash}p{#1}}
\newcolumntype{C}[1]{>{\centering\arraybackslash}p{#1}}
\newcolumntype{R}[1]{>{\raggedleft\arraybackslash}p{#1}}
\newcommand{\overbar}[1]{\mkern 1.5mu\overline{\mkern-1.5mu#1\mkern-1.5mu}\mkern 1.5mu}

\usepackage{algorithm}
\usepackage{algorithmic}
\usepackage{enumitem}
\setlist[itemize]{leftmargin=*}
\usepackage{verbatim} 

\lstdefinestyle{mystyle}{
    backgroundcolor=\color{backcolour},   
    commentstyle=\color{codegreen},
    keywordstyle=\color{magenta},
    numberstyle=\tiny\color{codegray},
    stringstyle=\color{codepurple},
    basicstyle=\ttfamily\footnotesize,
    breakatwhitespace=false,         
    breaklines=true,                 
    captionpos=b,                    
    keepspaces=true,                 
    numbers=left,                    
    numbersep=5pt,                  
    showspaces=false,                
    showstringspaces=false,
    showtabs=false,                  
    tabsize=2
}
 
\usepackage{scalerel,stackengine}
\stackMath
\newcommand\reallywidehat[1]{%
\savestack{\tmpbox}{\stretchto{%
  \scaleto{%
    \scalerel*[\widthof{\ensuremath{#1}}]{\kern-.6pt\bigwedge\kern-.6pt}%
    {\rule[-\textheight/2]{1ex}{\textheight}}
  }{\textheight}%
}{0.5ex}}%
\stackon[1pt]{#1}{\tmpbox}%
}

\title{Distributed deep reinforcement learning for simulation control}

\author{
  Suraj Pawar\thanks{This work was performed during a Research Aide appointment at the Argonne Leadership Computing Facility, Argonne National Laboratory}  \\
  School of Mechanical \& Aerospace Engineering,\\
  Oklahoma State University, \\
  Stillwater, Oklahoma - 74078, USA.\\
  \texttt{supawar@okstate.edu} \\
   \And
 Romit Maulik \\
  Argonne Leadership Computing Facility,\\
  Argonne National Laboratory, \\
  Lemont, Illinois, USA.\\
  \texttt{rmaulik@anl.gov} \\
}

\begin{document}
\maketitle

\begin{abstract}
Several applications in the scientific simulation of physical systems can be formulated as control/optimization problems. The computational models for such systems generally contain hyperparameters, which control solution fidelity and computational expense. The tuning of these parameters is non-trivial and the general approach is to manually `spot-check' for good combinations. This is because optimal hyperparameter configuration search becomes impractical when the parameter space is large and when they may vary dynamically. To address this issue, we present a framework based on deep reinforcement learning (RL) to train a deep neural network agent that controls a model solve by varying parameters dynamically. First, we validate our RL framework for the problem of controlling chaos in chaotic systems by dynamically changing the parameters of the system. Subsequently, we illustrate the capabilities of our framework for accelerating the convergence of a steady-state CFD solver by automatically adjusting the relaxation factors of discretized Navier-Stokes equations during run-time. The results indicate that the run-time control of the relaxation factors by the learned policy leads to a significant reduction in the number of iterations for convergence compared to the random selection of the relaxation factors. Our results point to potential benefits for learning adaptive hyperparameter learning strategies across different geometries and boundary conditions with implications for reduced computational campaign expenses.\footnote{Data and codes available at \url{https://github.com/Romit-Maulik/PAR-RL}}
\end{abstract}


\section{Introduction}
In recent years, there has been a proliferation in the use of machine learning (ML) for fluid mechanics problems that use the vast amount of data generated from experiments, field measurements, and large-scale high fidelity simulations \cite{brunton2019machine, brenner2019perspective}. \textcolor{blue}{The vast majority of ML applications for fluid mechanics have relied on supervised learning. Examples of frameworks that utilize this paradigm include artificial neural networks (ANN), Gaussian processes, and tree-based methods such as random forests. These methods, in particular ANNs, are popular due to their ability to approximate complex nonlinear functions from large amounts of data, although this comes at an expense of reduced interpretability. Supervised learning has been adopted for a variety of tasks such as turbulence closure modeling \cite{duraisamy2019turbulence},  reduced order modeling \cite{murata2020nonlinear}, super-resolution and forecasting  \cite{jiang2020meshfreeflownet,kim2020prediction}, solving partial differential equations \cite{raissi2019physics,long2019pde}, etc. However, a key limitation of supervised learning is that it relies on labeled data to learn relationships from observables to targets. Unfortunately, many practical engineering problems provide situations where labeled data is unavailable, thereby limiting the practicality of supervised ML.}



In contrast, reinforcement learning (RL) is another type of ML philosophy, where an agent interacts with an environment and learns a policy that will maximize a predefined long-term reward \cite{sutton2018reinforcement}. \textcolor{blue}{The long-term reward is generally constructed by human-designed inductive biases and does not require any specific labeled data. An example of a process reward relevant to fluid mechanics can be to maximize the negative of the drag constrained to a certain lift for flow around an airfoil. The RL deployment would then attempt to maximize this reward by searching the space of airfoil configurations for specific operating conditions \cite{rabault2019}.} Multiple methods can be employed as an agent in RL, but the most successful is an ANN in conjunction with RL (commonly referred as deep RL). Deep RL is gaining widespread popularity due to its success in robotics, playing Atari games, the game of GO, and process control \cite{silver2016mastering,spielberg2017deep}. 
Deep RL is also making inroads in fluid mechanics, as a means to solve fundamental problems in flow control. RL has been utilized to learn collective hydrodynamics of self-propelled swimmers to optimally achieve a specified goal \cite{verma2018efficient}, to train a glider to autonomously navigate in a turbulent atmosphere \cite{reddy2018glider}, for controlled gliding and perching of ellipsoid shaped bodies \cite{novati2019controlled}, and navigation of smart microswimmers in complex flows \cite{colabrese2017flow}. RL has also been used for closed-loop control of drag of a cylinder flow \cite{gueniat2016statistical}, flow control around bluff-bodies in experimental environments \cite{Fan26091}, control of two-dimensional convection-diffusion partial differential equation \cite{farahmand2017deep}, designing control strategies for active flow control \cite{rabault2019}, discovering computational models from the data \cite{bassenne2019computational}, and shape optimization \cite{ghraieb2020optimization}. RL's attractiveness stems from the ability to perform both exploitation and exploration to learn about a particular environment.

Many real-world system optimization tasks can be formulated as RL problems and offer a potential for the application of the recent advances in deep RL \cite{haj2019view}. Several problems in scientific simulations share a similarity with system optimization problems. For example, the iterative solver in numerical simulations of fluid flows may be considered as a plant to be controlled. The iterative solver consists of different parameters such as relaxation factors, smoothing operations, type of solver, residual tolerance, etc., and all these parameters could be controlled during run-time with RL to achieve faster convergence. Another example is the turbulence closure model employed in computational fluid dynamics (CFD) simulations to represent unresolved physics. The closure coefficients of the turbulence model can be controlled to provide the accurate turbulence statistics. Indeed, recently, multi-agent RL was introduced as an automated turbulence model discovery tool in the context of sub-grid scale closure modeling in large eddy simulation of homogeneous isotropic turbulence \cite{novati2020automating}. 

To illustrate the application of RL for the control of hyperparameters in computational models employed in scientific simulations, we consider the problem of accelerating the convergence of steady-state iterative solvers in CFD simulations by dynamically controlling relaxation factors of the discretized Navier-Stokes equations. Specifically, we train the RL agent to control relaxation factors to optimize the computational time or the number of iterations it takes for the solution to converge. This control problem differs from standard control problems, as the target state (i.e., the converged solution) is not known \textit{a priori}. There have been some studies done for dynamically updating relaxation factors with the goal to maximize the speed of convergence. One of the methods is to use a fuzzy control algorithm based on the oscillations of the solution norm observed over a large interval prior to the current iteration \cite{dragojlovic2004fuzzy}. Other methods based on neural network and fuzzy logic have also been proposed to automate the convergence of CFD simulations  \cite{dragojlovic2001tuning,ryoo2005control}. Some of the limitations of the fuzzy logic methods are that they are based on certain assumptions about the system that might not always be accurate, and they do not scale well to larger or complex systems. On the other hand, the neural network agent trained using RL can discover a more accurate decision-making policy to achieve a certain objective and can be applied for complex systems. The problem of controlling relaxation factors in CFD simulations is analogous to controlling the learning rate of stochastic gradient descent (SGD) algorithm and RL has been demonstrated to achieve better convergence of SGD than human-designed learning rate \cite{xu2017reinforcement}. In the present work, we introduce a novel method based on RL to accelerate the convergence of CFD simulations applied to turbulent flows and demonstrate the robustness of the method for the backward-facing step test case.

One of the major challenges in using RL for scientific simulations is the computational cost of simulating an environment \cite{verma2018efficient,rabault2019accelerating}. For example, the cost of CFD solver depends on several factors such as mesh refinement, complexity of the problem, level of required convergence, and can range between the order of minutes to order of several hours. RL is known to be less sample efficient and most popular RL strategies require a large number of interactions with the environment. Therefore, the RL framework should take the computational overhead of simulating an environment into account for scientific computing problems. One of the approaches to accelerate RL training is to run multiple environments concurrently on different processors and then collect their experience to train the agent. We refer to this multi-environment approach as distributed RL in this study. There are many open-source packages such as RLLib \cite{liang2018rllib}, Tensorforce \cite{schaarschmidt2017tensorforce}, SEED RL \cite{espeholt2019seed}, Acme \cite{hoffman2020acme} that can be exploited to achieve faster training with distributed RL. In addition to the distributed execution of the environment, the CFD simulations can also be parallelized to run on multiple processing elements. The list of contributions of this work can be summarized as follows:
\begin{itemize}
    \item A distributed deep RL framework is presented for the optimization and control of computational system parameters encountered in scientific simulations. 
    \item We apply the deep RL framework to the task of restoring chaos in transiently chaotic systems, and for accelerating the convergence of a steady-state CFD solver. 
    \item We highlight the importance of distributed RL and how it aids in accelerated training for scientific applications.
    \item We also provide an open-source code and set of tutorials for deploying the introduced framework on distributed multiprocessing systems.
\end{itemize}

This paper is organized as follows. Section~\ref{sec:rl} provides background information on RL problem formulation and present proximal policy optimization (PPO) algorithm \cite{schulman2017proximal}. In Section~\ref{sec:results}, we validate our framework using the test case of restoring chaos in chaotic systems. Then, we describe how controlling relaxation factors in CFD simulations converted to sequential decision-making problem and discuss the performance of an RL agent to generalize for different boundary conditions and different geometry setup. Finally, we conclude with the summary of our study and outline the future research directions in Section~\ref{sec:conclusion}. 

\section{Reinforcement Learning} \label{sec:rl}
\begin{figure}[htbp]
\centering
\mbox{\subfigure{\includegraphics[width=0.98\textwidth]{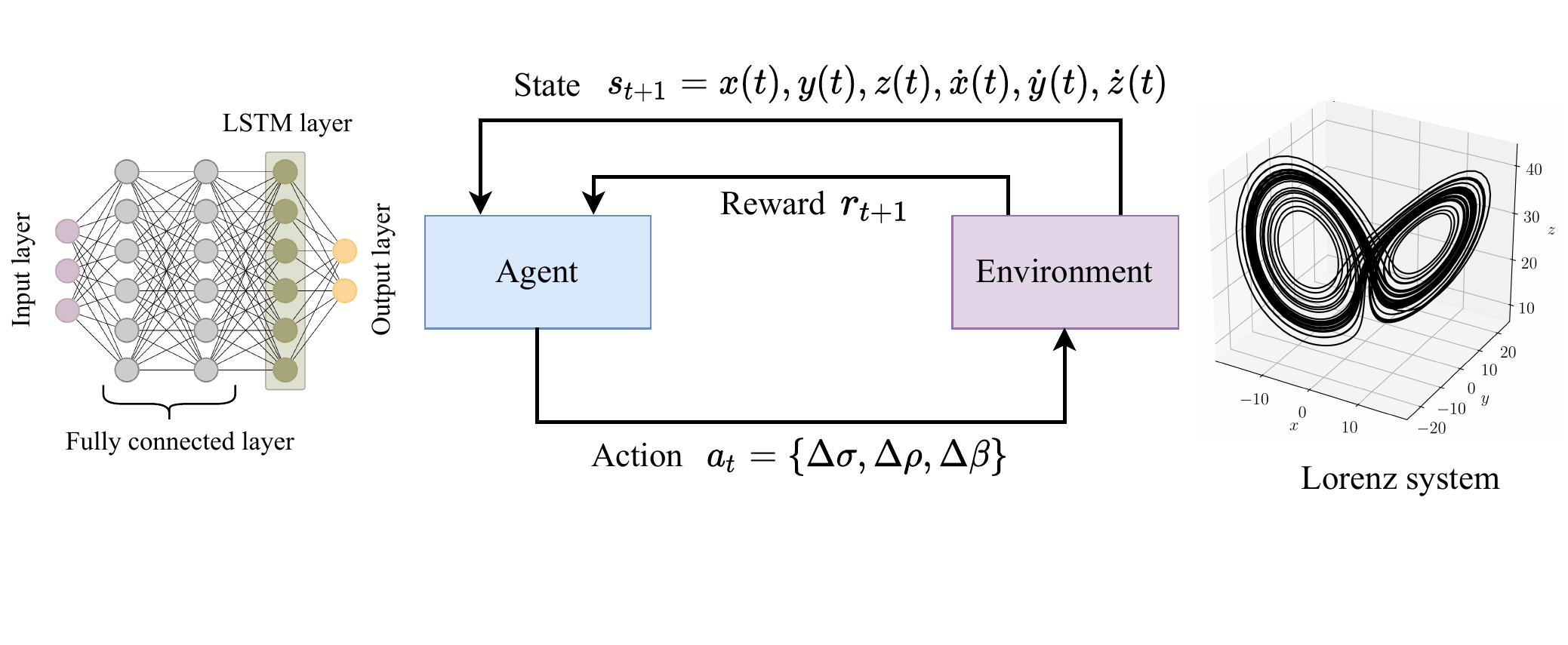}}}
\caption{\textcolor{blue}{Schematic of the deep RL framework for the chaotic process control example. The RL agent observes solution vector of the Lorenz system and its time-derivative as the state of the system and selects  perturbation to the control parameters as the action. The agent either receives positive or negative reward based on the magnitude of the velocity (Equations~\ref{eq:reward_lorenz_1} and ~\ref{eq:reward_lorenz_2}).  The neural network architecture is only for the purpose of representation.}}
\label{fig:rl_framework}
\end{figure}

In this section, we discuss the formulation of the RL problem and briefly describe the PPO algorithm \cite{schulman2017proximal} that belongs to a class of policy-gradient methods. In RL, the agent observes the state of the environment and then based on these observations takes an action. The objective of the agent is the maximization of the expected value of the cumulative sum of the reward. This problem statement can be framed as a Markov decision process (MDP). At each time step $t$, the agent observes some representation of the state of the system, $s_t \in \mathcal{S}$, and based on this observation selects an action, $a_t \in \mathcal{A}$. As a consequence of the action, the agent receives the reward, $r_t \in \mathcal{R}$ and the environment enters in a new state $s_{t+1}$. Therefore, the interaction of an agent with the environment gives rise to a trajectory as shown below
\begin{equation}
    \tau = \{ s_0,a_0,r_0,s_1,a_1,r_1,\dots \}.
\end{equation}
\textcolor{rev1}{Figure~\ref{fig:rl_framework} displays the schematic of deep RL framework for the chaotic process control example where the Lorenz system is utilized as an environment.} The advantage of the MDP framework is that it is flexible and can be applied to different problems in different ways. For example, the time steps in the MDP formulation need not refer to the fixed time step as used in CFD simulations; they can refer to arbitrary successive stages of taking the action. The goal of the RL agent is to choose an action that will maximize the expected discounted return over the trajectory $\tau$ and mathematically, it can be written as 
\begin{equation}
    {R}(\tau) = \sum_{t=0}^T \gamma^t r_t, \label{eq:reward}
\end{equation}
where $\gamma$ is a parameter called a discount rate and it lies between $[0,1]$, and $T$ is the horizon of the trajectory. The discount rate determines how much importance to be given to the long term reward compared to immediate reward. If $\gamma \rightarrow 0$, then the agent is concerned with maximizing immediate rewards and as $\gamma \rightarrow 1$, the agent takes future reward into account more strongly. The RL tasks are also classified based on whether they terminate or not. Some tasks involve agent-environment interactions that break into a sequence of episodes, where each episode ends in a state called the terminal state. On the other hand, there are tasks, such as process control, that that goes on continually without limit. The horizon $T$ of the trajectory for these continuing tasks goes to infinity in Equation~\ref{eq:reward}.

In RL, the agent's decision making procedure is characterized by a policy $\pi(s,a) \in {\Pi}$. The RL agent is trained to find a policy so as to optimize the expected return when starting in the state $s$ at time step $t$ and is called as V-value function. We can write the V-value function as 
\begin{equation}
    V^{\pi}(s) = \mathbb{E}_\pi \left[ \sum_{k=0}^\infty \gamma^k r_{t+k} | s_t=s, \pi \right].
\end{equation}
Similarly, the expected return starting in a state $s$, taking an action $a$, and thereafter following a policy $\pi$ is called as the Q-value function and can be written as 
\begin{equation}
    Q^{\pi}(s,a) = \mathbb{E}_\pi \left[ \sum_{k=0}^\infty \gamma^k r_{t+k} | s_t=s, a_t=a, \pi \right].
\end{equation}
We also define an advantage function that can be considered as an another version of Q-value function with lower variance by taking the V-value function as the baseline. The advantage function can be written as 
\begin{equation}
    A^{\pi}(s,a) = Q^\pi(s,a) - V^\pi(s).
\end{equation}

In deep RL, the neural network is used as an RL agent, and therefore, the weights and biases of the neural network are the parameters of the policy \cite{sutton2000policy}. We use $\pi_w(\cdot)$ to denote that the policy is parameterized by $w \in \mathbb{R}^d$. The agent is trained with an objective function defined as \cite{sutton2018reinforcement}
\begin{equation}
    J(w)~ \dot{=} ~ V^{\pi_w}(s_0), \label{eq:obj_function}
\end{equation}
where an episode starts in some particular state $s_0$, and $V^{\pi_w}$ is the value function for the policy $\pi_w$. The policy parameters $w$ are learned by estimating the gradient of an objective function and plugging it into a gradient ascent algorithm as follows
\begin{equation}
    w \leftarrow w + \alpha \nabla_w J(w),    \label{eq:gradient_ascent}
\end{equation}
where $\alpha$ is the learning rate of the optimization algorithm. The gradient of an objective function can be computed using the policy gradient theorem \cite{sutton2000policy} as follows
\begin{eqnarray}
    \nabla_w V^{\pi_w}(s_0) = \mathbb{E}_{\pi_w}\big[ \nabla_w \big(\log ~\pi_w(s,a) \big) Q^{\pi_w}(s,a)].
\end{eqnarray}

There are two main challenges in using the above empirical expectation. The first one is the large number of samples required and the second is the difficulty of obtaining stable and steady improvement. There are different families of policy-gradient algorithms that are proposed to tackle these challenges \cite{konda2000actor,schulman2015trust,schulman2015high}. The performance of policy gradient methods is highly sensitive to the learning rate $\alpha$ in Equation~\ref{eq:gradient_ascent}. If the learning rate is large it can cause the training to be unstable. The PPO algorithm uses a clipped surrogate objective function to avoid excessive update in policy parameters in a simplified way \cite{schulman2017proximal}. The clipped objective function of the PPO algorithm is 
\begin{equation}
    J^{{\rm clip}}(w) = \mathbb{E}\big[ {\rm min}(r_t(w) {A}^{\pi_w}(s,a), {\rm clip} (r_t(w),1-\epsilon,1+\epsilon) {A}^{{\pi_w}}(s,a) ) \big], \label{eq:ppo_clip}
\end{equation}
where $r_t(w)$ denotes the probability ratio between new and old policies as follow
\begin{equation}
    r_t(w) = \frac{\pi_{w+\Delta w}(s,a)}{\pi_{w}(s,a)}.
\end{equation}

The $\epsilon$ in Equation~\ref{eq:ppo_clip} is a hyperparameter that controls how much new policy is allowed to be deviated from the old. This is done using the function ${\rm clip}(r_t(w),1-\epsilon,1+\epsilon)$ that enforces the ratio between new and old policy ($r_t(w)$) to stay between the limit $[1-\epsilon,1+\epsilon]$. 

\section{Numerical Experiments} \label{sec:results}
In this section, we demonstrate the application of deep RL for two problems where the parameters of the systems need to be controlled to achieve a certain objective. The first test case is restoring chaos is chaotic dynamical systems \cite{vashishtha2020restoring}. We utilize this test case as a validation example for our distributed deep RL framework. The second test problem is accelerating the convergence of steady-state CFD solver through deep RL and we illustrate this for turbulent flow over a backward-facing step example.

\subsection{Chaotic process control}
The chaotic state is essential in many dynamical systems and the phenomenon called crisis can cause sudden destruction of chaotic attractors \cite{grebogi1983crises}. This can be negative in many applications where the chaos is essential such as to enhance mixing by chaotic advection \cite{ottino1990mixing} or to prevent the collapse of electric power systems \cite{dobson1989towards}. Some of the methods to sustain chaos is to use small perturbations based on knowledge of dynamical systems and \textit{a priori} information about the escape regions from chaos \cite{yang1995preserving}. Vashishtha et. al. \cite{vashishtha2020restoring} proposed an approach using deep RL to determine small perturbations to control parameters of the Lorenz system \cite{lorenz1963deterministic} in such a way that the chaotic dynamics is sustained despite the uncontrolled dynamics being transiently chaotic. This method does not require finding escape regions, target points, and therefore attractive for systems where the complete information about the dynamical system is unknown. 

\begin{figure}[htbp]
\centering
\mbox{\subfigure{\includegraphics[width=0.98\textwidth]{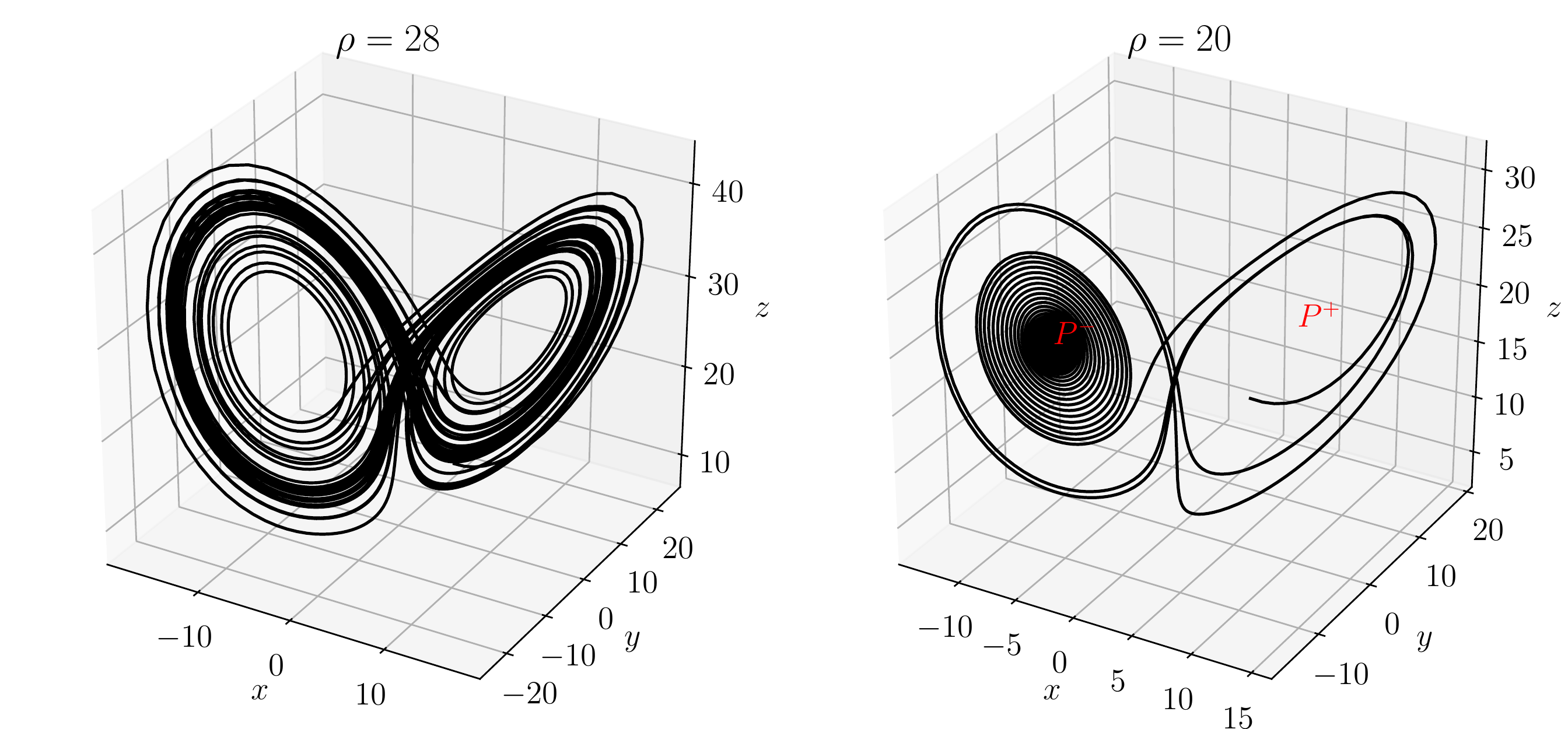}}}
\caption{Evolution trajectory of the Lorenz system with $\rho=28$ (left) and $\rho=20$ (right). }
\label{fig:l3}
\end{figure}

The Lorenz system is described by following equations 

\begin{eqnarray}
    \frac{dx}{dt} &= \sigma (y-x), \\
    \frac{dy}{dt} &= x (\rho - z) - y, \\
    \frac{dz}{dt} &= xy - \beta z, 
\end{eqnarray}
where $x,y,z$ are the state of the Lorenz system, and $\sigma,\rho,\beta$ are the system's parameter. The Lorenz system give rise to chaotic trajectory with $\sigma = 10$, $\rho = 28$, and $\beta = 8/3$  as shown in Figure~\ref{fig:l3}. However, the Lorenz system exhibits transient chaotic behavior for $\rho \in [13.93,24.06]$ \cite{kaplan1979preturbulence}. For example, if we use $\rho$ = 20 for the Lorenz system, the solution converges to a specific fixed point $P^-=(-7.12,-7.12,19)$ as illustrated in Figure~\ref{fig:l3}.  

In order to sustain the chaos in transiently chaotic Lorenz system, we utilize the similar RL problem formulation as Vashishtha et. al. \cite{vashishtha2020restoring}. The RL agent observes the solution vector and its time-derivative as the state of the system. Therefore, we have 
\begin{equation}
    s_t = x(t),y(t),z(t),\dot{x}(t), \dot{y}(t),\dot{z}(t),
\end{equation}
where the dot represent the time-derivative, i.e., velocity. Based on this observations, the agent choose an action which is the perturbation to the control parameters of the Lorenz system as given below
\begin{equation}
    a_t = \{ \Delta \sigma, \Delta \rho, \Delta \beta \}.
\end{equation}
The perturbation to the control parameters is restricted to lie within a certain interval which is $\pm$ 10\% of its values. Therefore, we have $\Delta \sigma \in [-\sigma/10, \sigma/10]$, $\Delta \rho \in [-\rho/10, \rho/10]$, and $\Delta \beta \in [-\beta/10, \beta/10]$. We note here that these limits remain fixed throughout the episode based on the initial values of $(\sigma,\rho,\beta)$. For a transiently chaotic system, the trajectory converges to a fixed point with time and therefore the magnitude of velocity will decrease consistently with time. This fact is utilized to design the reward function. When the magnitude of velocity is greater than certain threshold velocity $V_{{\rm Th}}$, the agent is rewarded and the agent is penalized when the magnitude of velocity is less than $V_{{\rm Th}}$. In addition to assigning reward for each time step, a terminal reward is given at the end of each episode. The reward function can be written as 
\begin{align}\label{eq:reward_lorenz_1}
    r_t &= \begin{cases}
    10, & V(t) > V_{\text{Th}}\\
    -10, & V(t) \le V_{\text{Th}}
    \end{cases}, \\
    r_{\text{terminal}} &= \begin{cases} \label{eq:reward_lorenz_2}
    -100, & \overbar{r_t} \le -2\\
    0, & \overbar{r_t} > -2
    \end{cases},
\end{align}
where $\overbar{r_t}$ is the average of stepwise reward over the last 2000 time steps of an episode. The value of the threshold velocity $V_{{\rm Th}}$ is set at 40. 

For training the agent, we divide each episode into 4000 time steps with size $dt = 2 \times 10^{-2}$. We train the agent for $2 \times 10^5$ time steps which corresponds to 50 episodes. The RL agent is a fully connected neural network with two hidden layers and 64 neurons in each layer. Additionally, the output of the second hidden layer is further processed by the LSTM layer composed of 256 cells. The hyperbolic tangent (tanh) activation function is used for all hidden layers. \textcolor{rev1}{We utilize the same architecture as the one used in the original study \cite{vashishtha2020restoring}. From the point of view of optimal agent architecture selection, there are methods like population based training that have been applied for optimizing the parameters of deep RL and hyperparameters of the agent \cite{jaderberg2017population}. However, considering the episode hungry nature of RL and computational expense of the environment, the hyperparameter search of the agent in deep RL can quickly become computationally intractable.}

\begin{figure}[htbp]
\centering
\mbox{\subfigure{\includegraphics[width=0.98\textwidth]{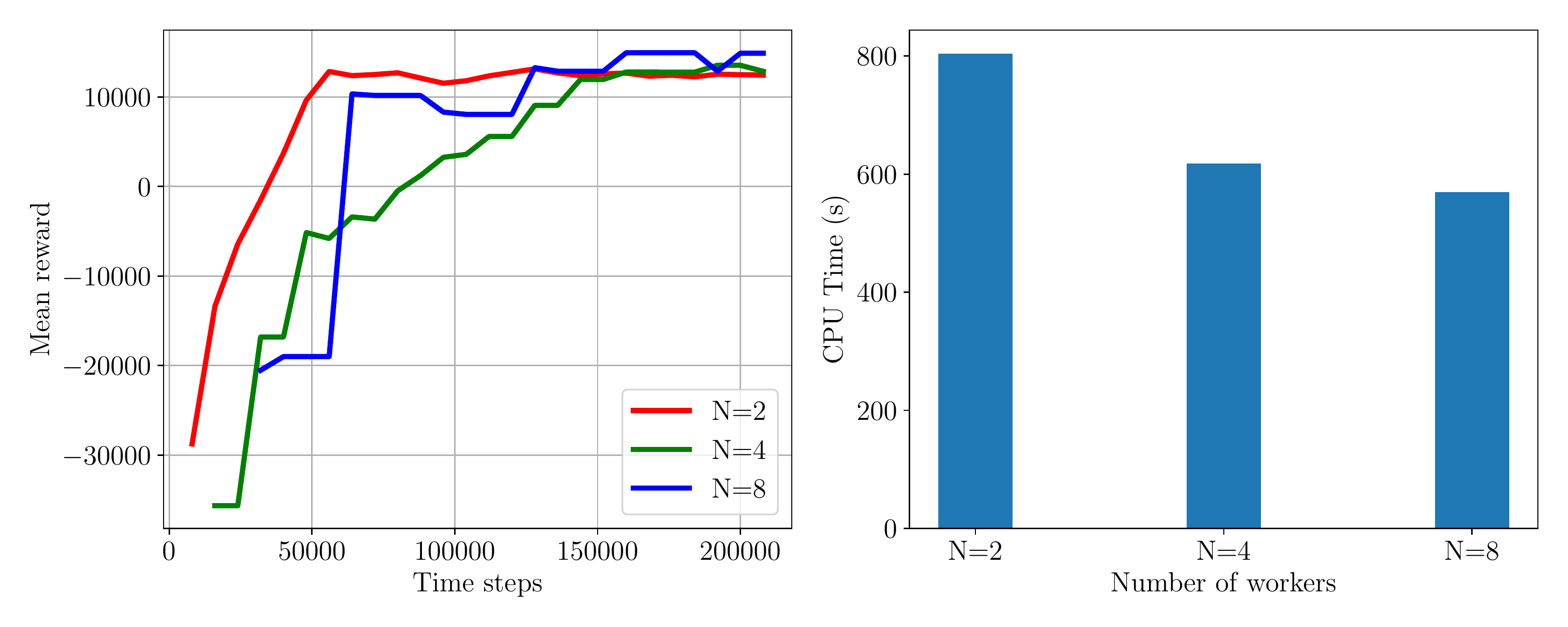}}}
\caption{Evolution of the moving average mean reward for different number of workers trained with the PPO algorithm (left). The moving average window size is set at 5 episodes. The CPU time required for training an agent with different number of workers (right).}
\label{fig:mean_reward_lorenz}
\end{figure}

Figure~\ref{fig:mean_reward_lorenz} displays how the mean reward is progressing with training for a different number of workers. We note here that by a different number of workers, we mean that the agent-environment interactions are run concurrently on different processors. In distributed deep RL, the environment is simulated on different processors and their experience is collected. These sample batches encode one or more fragments of a trajectory and are concatenated into a larger batch that is then utilized to train the neural network. More concretely, in this particular example, we collect the experience of an agent for 8000 time steps to train the neural network. When we employ two workers, the agent-environment interaction is simulated for 4000 time steps on each worker and then this experience is concatenated to form a training batch for the neural network. Similarly, for four workers, the agent-environment interaction is simulated for 2000 time steps on each worker to collect a training batch size of 8000 time steps. Figure~\ref{fig:mean_reward_lorenz} also reports the computational time required for training. As we increase the number of workers from two to four, the computational speed is approximately reduced by 25\%. However, an increase in the number of workers to eight leads to a marginal improvement in terms of CPU time. This is not surprising, as the communication frequency is increased with eight workers and hence there is no significant reduction in CPU time.

\begin{figure}[htbp]
\centering
\mbox{\subfigure{\includegraphics[width=0.98\textwidth]{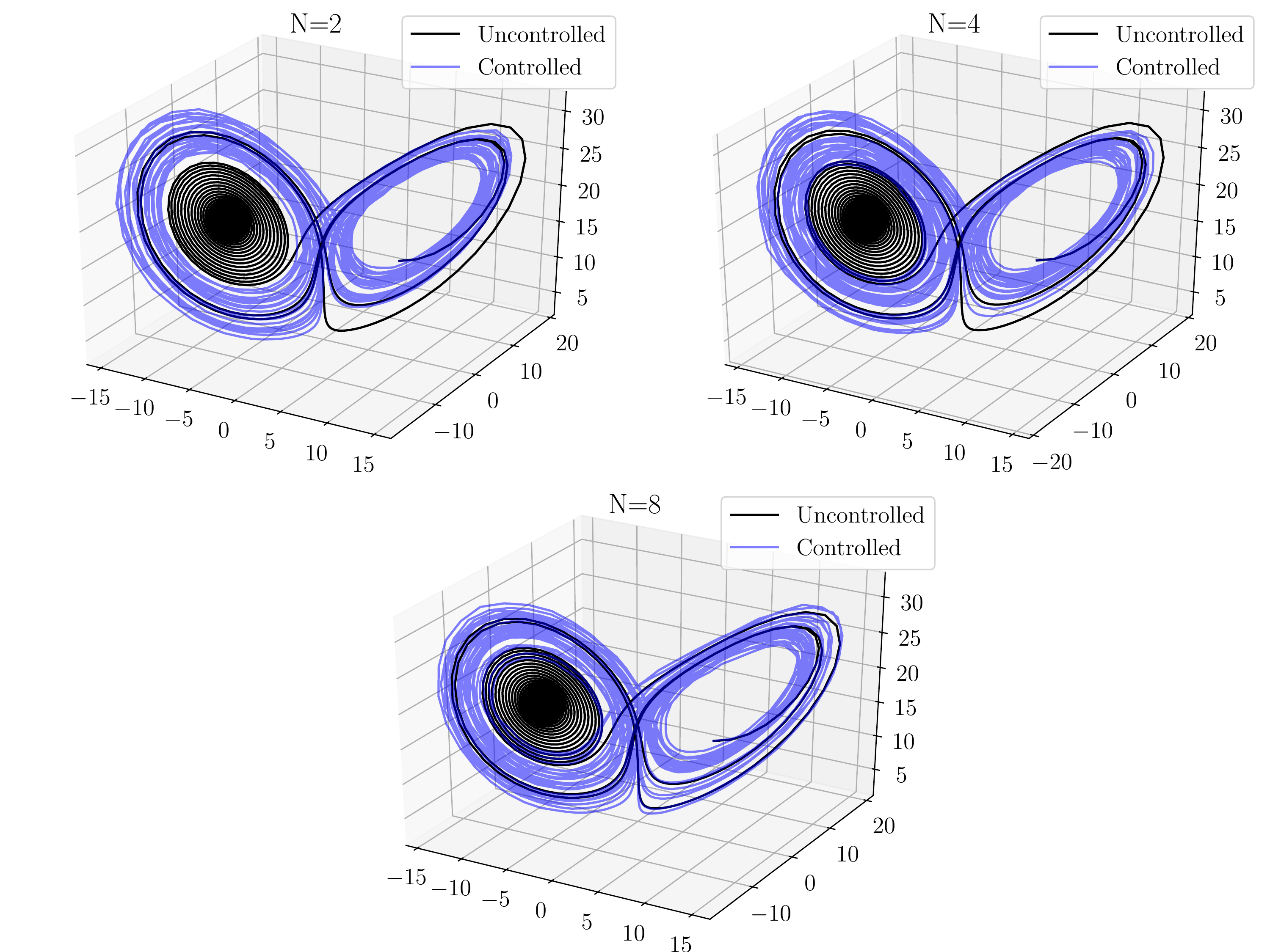}}}
\caption{Comparison of the evolution of the Lorenz system without applying the control (black) and with an application of the control (blue) policy learned by an RL agent trained using different number of workers (i.e., $N=2,4,8$). The parameters of the Lorenz system are $(\sigma,\beta,\rho)=(10,8/3,20)$ and the initial condition is $(x,y,z)=(1,12,9)$.}
\label{fig:results_lorenz}
\end{figure}

Figure~\ref{fig:results_lorenz} depicts the performance of the trained RL agent in sustaining chaos in a transiently chaotic Lorenz system. In Figure~\ref{fig:results_lorenz}, the black trajectory is for uncontrolled solution of the Lorenz system for parameters $(\sigma,\rho,\beta)=(10,20,8/3)$. When the policy learned by an RL agent is used for the control, the Lorenz system does not converge to a fixed point of $P^-$. We can also see that the RL agent trained with a different number of workers can learn the policy effectively and is able to sustain chaos over a period of temporal integration. This observation is very important for employing RL in physical systems where the simulation of a computationally intensive environment is the bottleneck and distributed deep RL can be employed to address this issue.   

\subsection{Solver convergence acceleration}
In CFD, the Navier-Stokes equations are solved by discretizing them which leads to a system of linear equations. This system of linear equations is usually solved using iterative techniques like Gauss-Seidel methods, Krylov subspace methods, multigrid methods, etc. \cite{saad2003iterative}. The main logic behind iterative methods is to start with some approximate solution and then reduce the residual vector at each iteration, called relaxation steps. The relaxation steps are performed until the convergence criteria are reached. In the relaxation step, the value of the variable at the next time step is obtained by adding some fraction of the difference between the current value and predicted value to the value of the variable at the current time step. The SIMPLE algorithm \cite{patankar2018numerical} is widely used in CFD and it utilizes the underrelaxation method to update the solution from one iteration to another. The underrelaxation methods improve the convergence by slowing down the update to the solution field. The value of the variable at the $k^{{\rm th}}$ iteration is obtained as the linear interpolation between the value from the previous iteration and the value obtained in the current iteration as follows

\begin{equation}
    x_P^{(k)} = x_P^{(k-1)} + \alpha \bigg( \frac{\sum a_{{\rm nb}} x_{{\rm nb}} + b}{a_P} - x_P^{(k-1)} \bigg), \label{eq:simple}
\end{equation}
where $0 < \alpha < 1$ is the underrelaxation factor, $x_P^{(k)}$ is the value of the variable at node $P$ to be used for the next iteration, $x_P^{(k-1)}$ is the value of the variable at node $P$ from the previous iteration, $x_{{\rm nb}}$ is the values of the variables at the neighboring nodes, and $a_P,~a_{{\rm nb}},~b$ are the constants obtained by discretizing Navier-Stokes equations. The underrelaxation factor $\alpha$ in Equation~\ref{eq:simple} should be chosen in such a way that it is small enough to ensure stable computation and large enough to move the iterative process forward quickly. In practice, a constant value of the relaxation factor is employed throughout the computation and this value is usually decided in an ad-hoc manner. Also, the suitable value of the relaxation factor is problem-dependent and a small change in it can result in a large difference in the number of iterations needed to obtain the converged solution. In the following we attempt to obtain a dynamic adaptation of the underrelaxation factor using deep RL.

\begin{figure}[htbp]
\centering
\mbox{\subfigure{\includegraphics[width=0.95\textwidth]{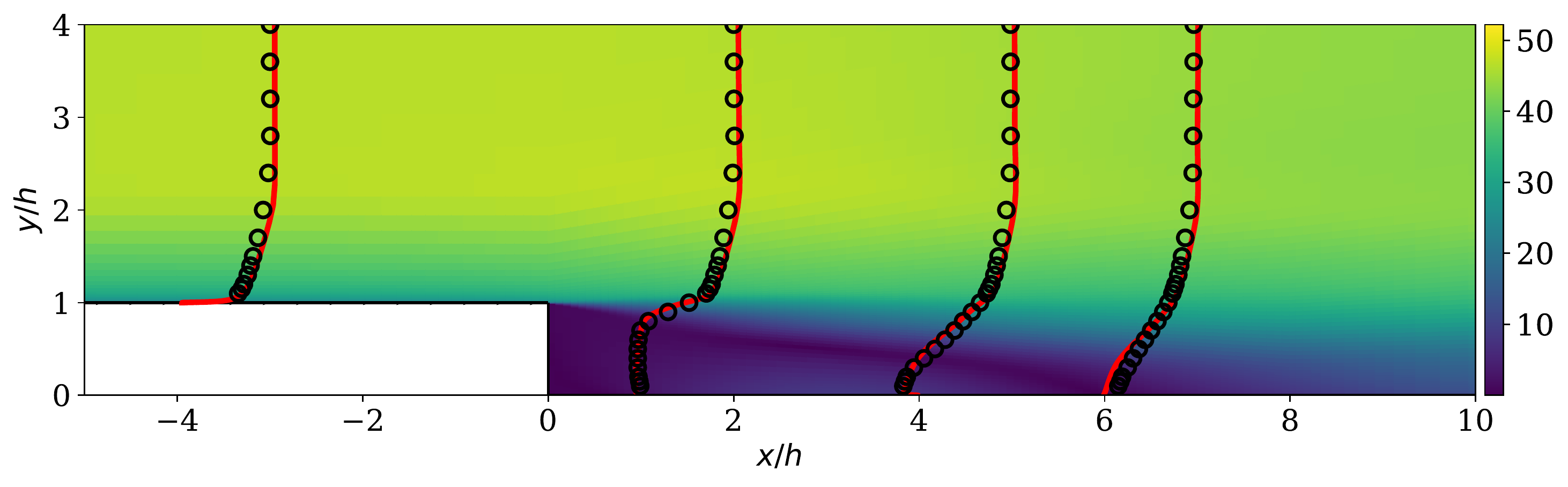}}}
\caption{Validation of the CFD model with experimental measurements for normalized velocity in the $x$-direction at four different axial locations. The experimental data for velocity measurements are published in \cite{driver1985features}. \textcolor{blue}{The contours indicate the velocity magnitude of the flow field.}}
\label{fig:velocity_profile}
\end{figure}

\begin{figure}[htbp]
\centering
\mbox{\subfigure{\includegraphics[width=0.95\textwidth]{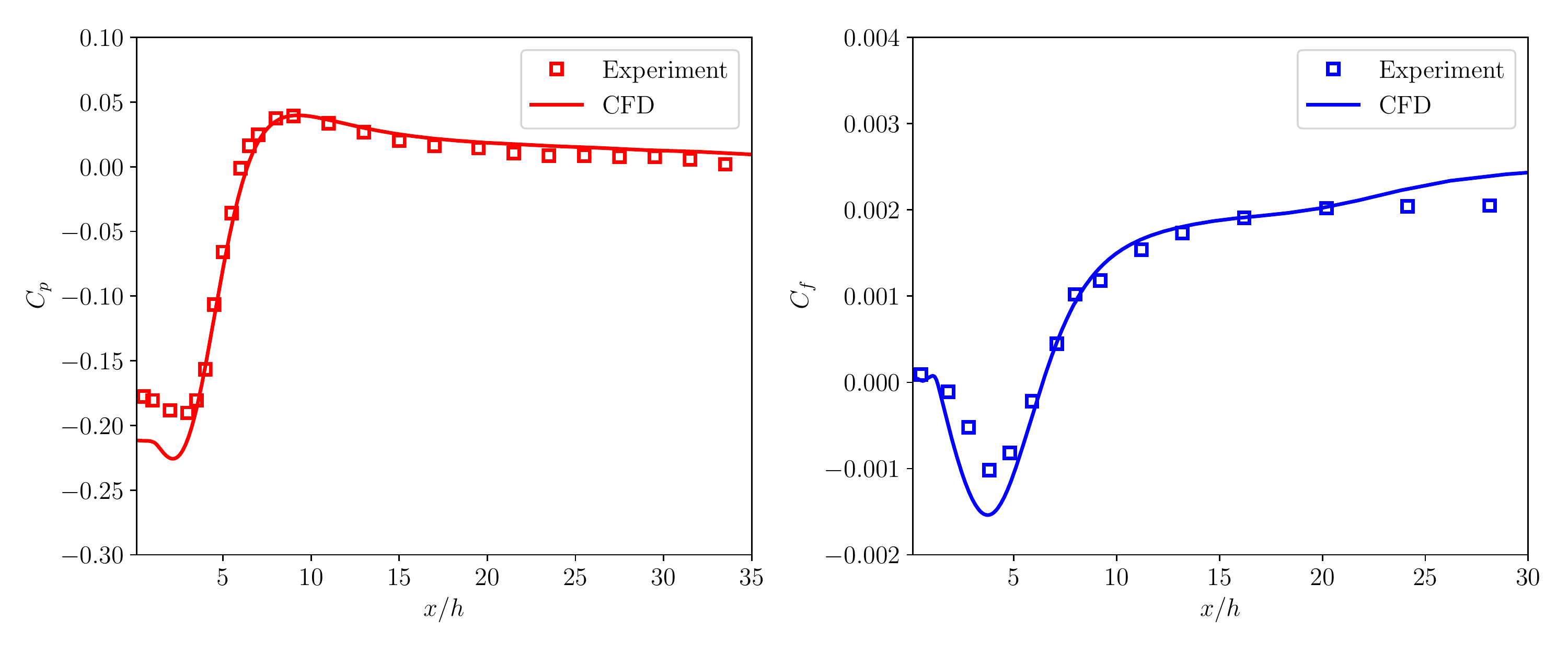}}}
\caption{Comparison of the pressure coefficient (left) and skin-friction coefficient (right) at the lower wall predicted by CFD simulation and experimental measurements. The experimental data for these coefficients are published in \cite{driver1985features}.}
\label{fig:cp_cf}
\end{figure}

\textcolor{blue}{We choose a CFD model for a two-dimensional canonical backward facing step that is commonly used for solver and turbulence model validation. The height of the step is $H=0.0127$ m and the free-stream reference velocity (flowing left to right) is $U_{{\rm ref}}=44.2$ m/s. The corresponding Reynolds number based on step height and reference free-stream velocity is approximately ${\rm Re}_H=36,000$. The turbulent intensity at the inlet is $I_t=6.1 \times 10^{-4}$ and the eddy-viscosity-to-molecular-viscosity ratio at the inlet is $\mu_t/\mu=0.009$. First, we validate a CFD model corresponding to the aforementioned geometry and flow conditions using the experimental data provided in \cite{driver1985features}. We use the computational mesh distributed by NASA \cite{rumsey2015recent} for our study. We utilize the steady-state simpleFoam solver in OpenFoam \cite{jasak2007openfoam} and apply the $k-\omega$ SST turbulent model \cite{menter2003ten} with the standard values of closure coefficients to simulate the turbulent flow. Figure~\ref{fig:velocity_profile} displays the contour plot of the magnitude of the velocity along with the normalized stream-wise velocity at four different axial locations obtained experimentally and numerically. We can observe that there is a very good match between the experimental results and numerical results for normalized stream-wise velocity.} \textcolor{rev1}{As shown in Figure~\ref{fig:cp_cf}, we can also see that there is a very good agreement between the experimental measurements and CFD prediction for the pressure coefficient and skin-friction coefficient on the lower wall.}

\begin{figure}[htbp]
\centering
\mbox{\subfigure{\includegraphics[width=0.98\textwidth]{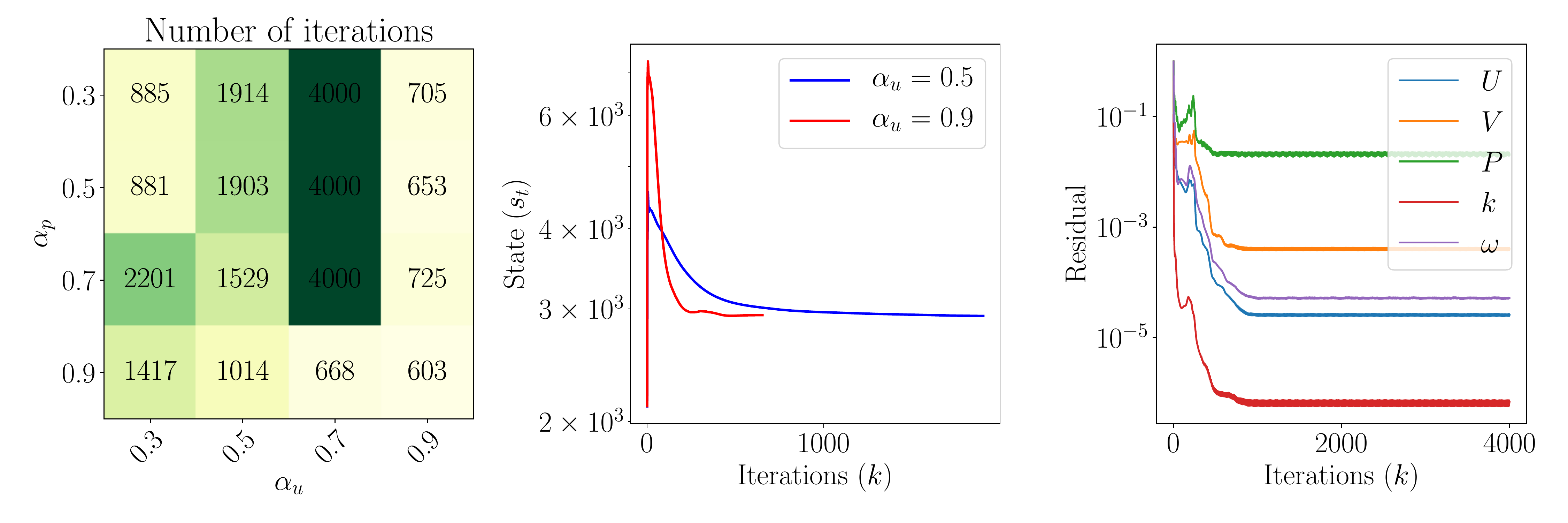}}}
\caption{\textcolor{blue}{Number of iterations required for the convergence with different combinations of relaxation factor for momentum and pressure equation (left). The convergence of the state of an environment with iterations for two values of momentum equation relaxation factor (middle). The pressure relaxation factor is set at $\alpha_p=0.5$ for both cases. Typical residual history for momentum equation, pressure equation, and turbulent quantities equation (right) for the backward facing step test case.}}
\label{fig:heatmap}
\end{figure}

\textcolor{blue}{Subsequently, we perform a manual analysis of the convergence characteristics of the CFD model for various underrelaxation factors. In Figure~\ref{fig:heatmap}, the number of iterations required for convergence with different values of relaxation factors for momentum and pressure equation is shown for the problem investigated in this study (i.e., the backward-facing step example). \textcolor{rev1}{Typical residual histories for different variables that are solved in the simulation of the backward-facing step example are also displayed in Figure~\ref{fig:heatmap}. We perform a trial simulation to assess convergence by running a simulation for 4000 iterations with $\alpha_u=0.9$ and $\alpha_p=0.9$. Based on this residual history, we can assume that the convergence is reached once the residual for momentum equation and turbulent quantities equation falls below $5 \times 10^{-4}$, and the residual for pressure equation is reduced below $5 \times 10^{-2}$.} It can be noticed that the number of iterations is less for $\alpha_u=0.9$ (momentum equation relaxation factor), for all values of $\alpha_p$ (pressure equation relaxation factor). However, for $\alpha_u=0.7$, the solution converges only for $\alpha_p=0.9$, and for other values of $\alpha_p$ the solution does not converge within 4000 iterations. Even though for most of the CFD simulations, a good value for relaxation factors can be found through trial and error, the process can become intractable for large parameter space. The complexity increases even further when the relaxation factor needs to be updated on the fly during run-time for multiphysics problems. Therefore, we attempt to automate the procedure of dynamic update of the relaxation factors using deep RL to accelerate the convergence of CFD simulations.} 

To learn a policy that is generalizable for different inlet boundary condition, we train the RL agent for multiple values of inlet velocities. For each episode, the inlet velocity is selected randomly from the interval between [35,65] m/s. We reiterate here that, in our study, the frequency at which the relaxation factors are updated is different from the frequency of CFD iterations. In particular, we update the relaxation factor for the momentum and pressure equation after every 100 iterations of the CFD simulation. \textcolor{rev1}{One of the advantages of supervised learning methods such as physics-informed neural network is that it allows us to embed the knowledge of physics explicitly in the learning process \cite{raissi2019physics,Taghizadeh_2020}. In contrast to supervised learning, the only opportunity for physics representation in the RL paradigm is the choice of the state, where the user has to design an appropriate quantity representative of the physics that balances expressiveness and conciseness. Therefore, the state of the system should be appropriately designed based on the physics of the problem, and the maximum information is made available to the RL agent while being simple in construction.} We use the average value of the square of the velocity magnitude as the characteristic quantity that represents the CFD solution. Thus, the state of the environment can be written as  
\begin{equation}
    s_t = \frac{1}{N_b}\sum_{i=1}^{N_b} [U_b(i)]^2 + \frac{1}{N_m}\sum_{i=1}^{N_m} [U_m(i)]^2,
\end{equation}
where $U_b$ is the velocity at the inlet boundary, $U_m$ is the nodal value of the velocity at internal mesh, $N_b$ is the total number of nodes at inlet boundary, and $N_m$ is the total number of nodes in the internal mesh. At the start of an episode, the velocity in the interior of the domain is initialized with zero velocity, and hence the initial state for each episode will depend only on the value of the velocity at the inlet boundary. Figure~\ref{fig:heatmap} displays how the state of the environment progress towards the converged state for two different values of relaxation factor for momentum equations. As the solution starts converging, the change in the state of the system from one iteration to other will also be reduced. \textcolor{rev1}{We note here that some other quantity based on the pressure as the characteristic quantity can also be utilized for the state of the system. Indeed, one can also use the iterative oscillations of the characteristic quantity observed over a large interval of the consecutive iterations before the current iteration as the state of the system. The information about fluctuating quantities may improve the effectiveness of the deep RL approach and we plan to consider them in our future studies.}

Based on the observation of the state of the system, the agent decides the relaxation factor for the momentum and pressure equations. The action of the agent can be written as  
\begin{equation}
    a_t = \{\alpha_u, \alpha_p \},
\end{equation}
where $\alpha_u$ and $\alpha_p$ are the relaxation factors for momentum and pressure equation, respectively. The action space for both relaxation factors lies within the interval $[0.2,0.95]$. Even though the action space is composed of only two relaxation factors, this problem is complicated due to its dynamic nature and simple exploratory computation will not be feasible. For more complex multiphysics problems, the action space can be easily augmented in this framework to include relaxation factors for other governing equations. The goal of the RL agent is to minimize the number of iterations required for obtaining the converged solution. We give the total number of iterations of the CFD simulation between two time steps as the reward for that time step. Accordingly, the reward at each time step can be written as  
\begin{equation}
    r_t = -(K_t - K_{t-1}), \label{eq:reward_solver}
\end{equation}
where $K$ is the iteration count of CFD simulation. Since each episode comes to end with the terminal state, the task of accelerating CFD simulations can be considered as an episodic task. Therefore, we set the discount factor $\gamma = 1.0$ for this test case. The stepwise reward function computed using Equation~\ref{eq:reward_solver} and with $\gamma = 1.0$ will give cumulative reward equal to the total number of iterations it took for CFD simulation to converge. Specifically, if the CFD simulation required 450 iterations to converge and the relaxation factor is updated every 100 iterations, then the trajectory for the reward will look like $\{-100,-100,-100,-100,-50\}$. 


For this example, the RL agent is a fully connected neural network with 64 neurons in each hidden layer. The RL agent is trained for 2000 episodes and the learning rate of an optimizer is set to $2.5\times 10^{-4}$. \textcolor{rev1}{For training an RL agent,} we assume that the convergence for CFD simulation is reached once the residual for momentum equation \textcolor{rev1}{and turbulent quantities} falls below $1 \times 10^{-3}$ and the residual for pressure equation is reduced below $5 \times 10^{-2}$. \textcolor{rev1}{We highlight here that during the testing phase of an agent, the CFD simulation is run till the residual for momentum equation and turbulent quantities drop below $5 \times 10^{-4}$. The use of the slightly mild criteria for the convergence of the CFD simulation during training allows us to train an agent in a computationally efficient manner. Indeed in practical CFD simulations, it is possible to ease on underrelaxation factors as we reach higher time steps to accelerate the convergence of CFD simulation.}

\begin{figure}[htbp]
\centering
\mbox{\subfigure{\includegraphics[width=0.95\textwidth]{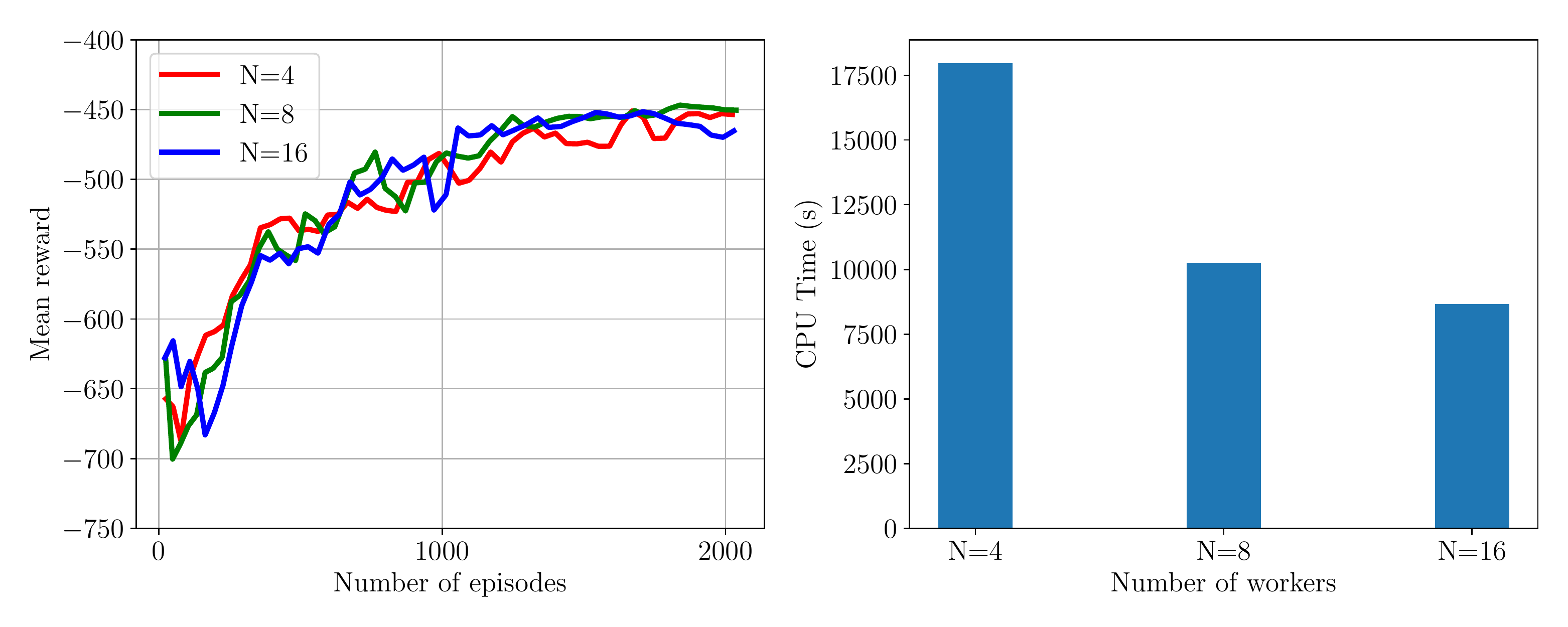}}}
\caption{Evolution of the moving average mean reward for the PPO algorithm trained using different number of workers (left). The moving average window size is set at 100 episodes. The CPU wall time for agent's training for different number of workers (right)}
\label{fig:mean_reward_solver}
\end{figure}

Figure~\ref{fig:mean_reward_solver} shows how the moving average mean reward progress with training for a different number of workers. The moving average window size is set at 100 episodes, i.e., the average of the reward is taken over the last 100 episodes. At the beginning of the training, the agent is dominantly exploring and the mean reward at the start of training is in the range of 650-700 iterations for CFD simulation to converge. The mean reward improves over the training and the final mean reward is around 450 iterations for the convergence of CFD simulations. This corresponds to approximately 30\% improvement in the number of iterations required for convergence. We highlight here that, for many CFD problems, suitable values of relaxation factors can be chosen to achieve faster convergence based on prior experience. However, for complex systems, the process of learning a rule to dynamically update relaxation factors through exploratory computations can become unmanageable. For such systems, the proposed RL framework can be attractive to achieve acceleration in the convergence. In Figure~\ref{fig:mean_reward_solver}, the CPU time for training is reported for a different number of workers. As we increase the number of workers/environments from 4 to 8, we get around 40\% improvement in the CPU time for the training. The improvement in CPU speed is only marginal as the number of processor is increased to 16, which might be due to increased communication between the processors. We note here that our environment, i.e. the CFD simulation utilizes a single core as the backward-facing step is a relatively small problem and can be simulated within an order of seconds or minutes depending upon the computer architecture. For high-dimensional problems, the parallelization through domain-decomposition can also be exploited to reduce the CPU time of each CFD simulation run.  

\begin{figure}[htbp]
\centering
\mbox{\subfigure{\includegraphics[width=0.95\textwidth]{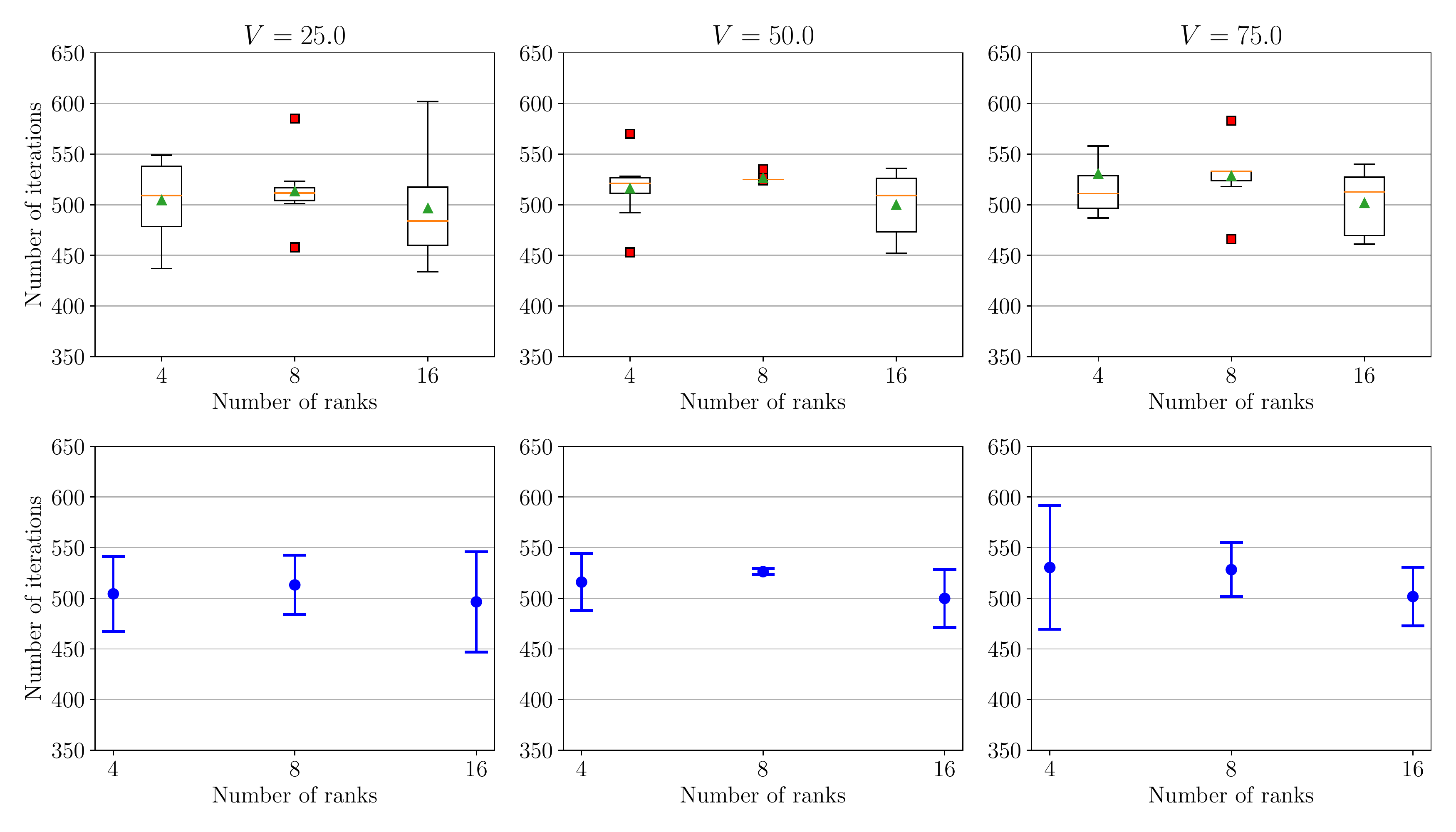}}}
\caption{Evaluation of the trained agent's PPO policy for different inlet velocities - $V=25.0$ m/s (left), $V=50.0$ m/s (middle), $V=75.0$ m/s (right). The top plot is the box plot in which the green marker is for the mean, red marker is for outliers, and the orange line is for the median of the data. The bottom plot shows the mean of the sample with error bars denoting the standard deviation of samples. }
\label{fig:results_train}
\end{figure}

\begin{figure}[htbp]
\centering
\mbox{\subfigure{\includegraphics[width=0.95\textwidth]{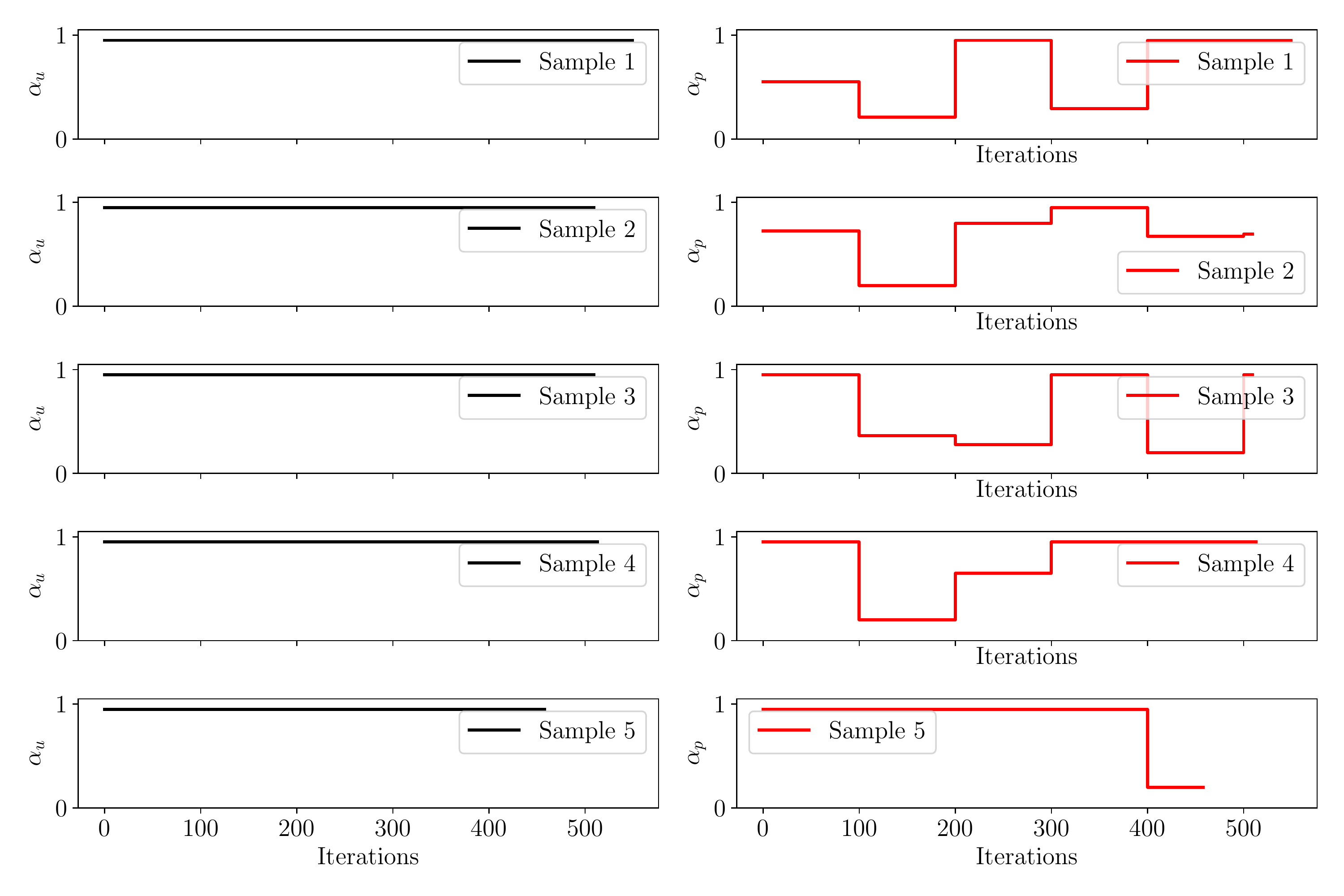}}}
\caption{Variation of the relaxation factor for the momentum equation and pressure equation over iterations of the steady-state CFD solver for $V=25.0$ m/s for five samples evaluated during testing with the learned policy.}
\label{fig:alpha_u_p}
\end{figure}

Once the agent is trained, we test the agent's performance for three different values of inlet velocity, $V=25.0,~50.0,~75.0$. Among these three velocities, the velocity $V=50.0$ lies within the training velocity range, i.e., [35.0, 65.0]. We run ten different samples of CFD simulation for these three inlet velocities, and utilize the RL agent to select the relaxation factor after every 100 iterations. Figure~\ref{fig:results_train} shows the box plot to depict the data obtained from ten samples through their quartiles. It should be noted that the policy learned by an agent is not deterministic, and hence the number iterations required for convergence for a single inlet velocity is not constant for all samples. The mean and standard deviation of these samples are also shown in Figure~\ref{fig:results_train}. From both these plots, we can see that the mean of the samples for all three different velocities is around 500 iterations. Therefore, we can conclude that the policy learned by the RL is generalizable to inlet boundary condition different than the training boundary condition. We plot the variation of relaxation factor for momentum equation and pressure equation in Figure~\ref{fig:alpha_u_p} to understand the policy learned by an RL agent. The RL agent has learned to keep the relaxation factor for momentum equation at its highest value (i.e., 0.95) throughout the CFD simulation. We do not see any specific pattern for the variation of pressure equation relaxation factor. Since, the relaxation factor for momentum equations is constant, the difference in number of iterations to convergence for different samples is mainly due to how the pressure relaxation factor is decided by an RL agent at various stages of the CFD simulation. \textcolor{rev1}{This behavior of the relaxation factors may be caused by the underlying nature of the Poisson pressure and momentum governing equations. However, there are multiple factors that engender this behavior aside from the governing equations such as the quality of the mesh and the discretization schemes used for the pressure Poisson solver and the momentum equations. At this point, we can not explicitly establish that causal link here despite the evidence of correlation.} 

\begin{figure}[htbp]
\centering
\mbox{\subfigure{\includegraphics[width=0.9\textwidth]{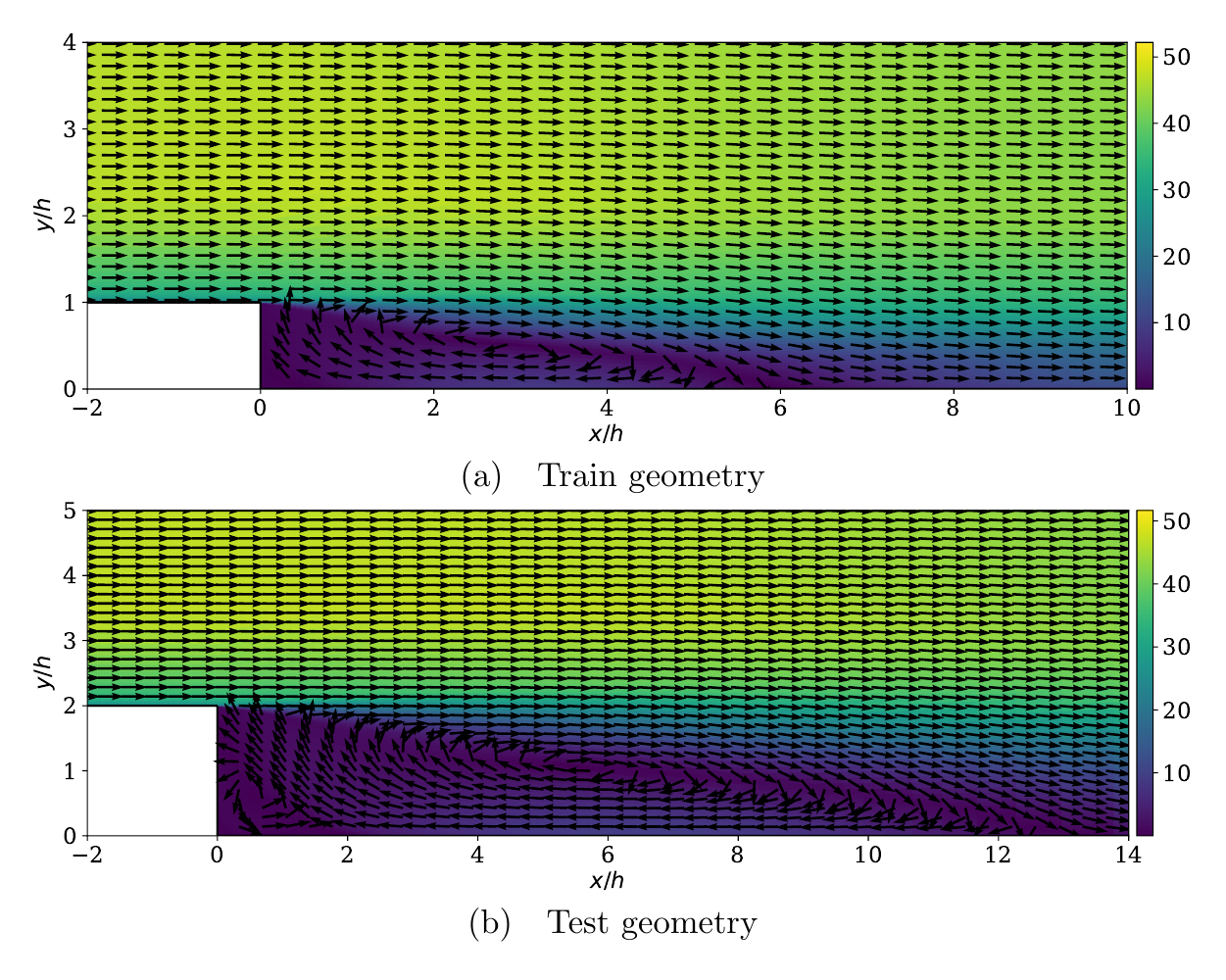}}}
\caption{Flow features in the vicinity of backward facing step for train and test geometry setup. The agent is trained with a step height $h=0.0127$ m and the test geometry has step height $h=0.0254$ m .}
\label{fig:contour_train_test}
\end{figure}

\begin{figure}[htbp]
\centering
\mbox{\subfigure{\includegraphics[width=0.95\textwidth]{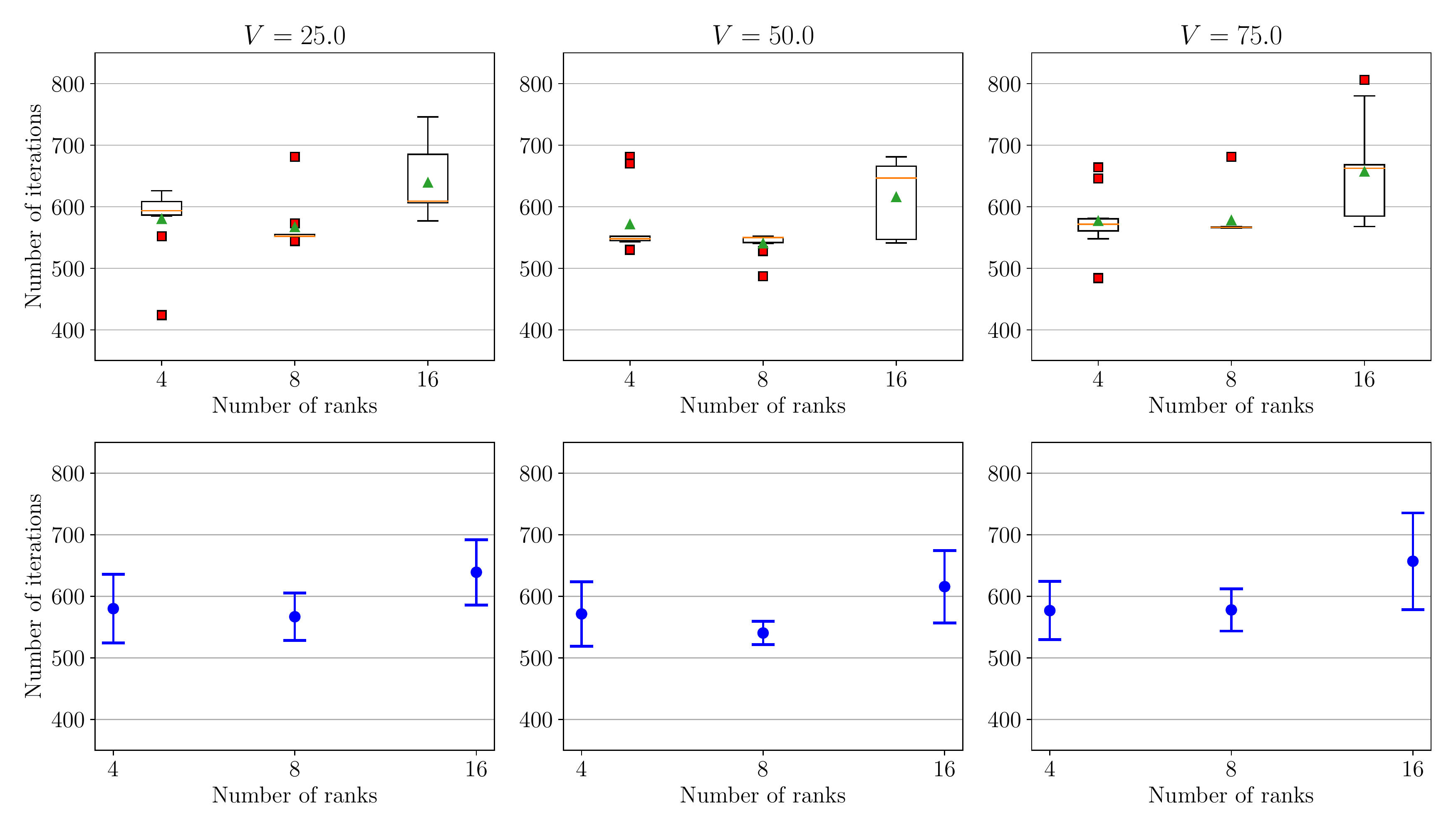}}}
\caption{Evaluation of the trained agent's PPO policy for different inlet velocities in test geometry setup - $V=25.0$ m/s (left), $V=50.0$ m/s (middle), $V=75.0$ m/s (right). The top plot is the box plot in which the green marker is for the mean, red marker is for outliers, and the orange line is for the median of the data. The bottom plot shows the mean of the sample with error bars denoting the standard deviation of samples.}
\label{fig:results_test}
\end{figure}

Furthermore, we test the learned policy for a different backward-facing step geometry. The test geometry has the step height twice that of the train geometry. In Figure~\ref{fig:contour_train_test}, the flow feature for train and test geometry is shown, and it can be seen that there is a sufficient difference in terms of the flow in the recirculation region. We run a similar experiment with the test geometry for three different values of inlet velocity. Figure~\ref{fig:results_test} displays the summary of the RL agent's performance in accelerating CFD simulation of the test geometry. Overall, the number of iterations required for the convergence for the test geometry is higher than the train geometry. The mean of the samples for test geometry is around 600 iterations for all three velocities. Interestingly, the variance of test geometry samples employed with the policy trained using 8 workers is less compared to the policy trained using 4 and 16 workers. 

\textcolor{rev1}{We emphasize here that the computational efficiency of the deep RL is one of the major bottlenecks compared to other methods such as PDE-constrained adjoint optimization. There have been several works done on applying adjoint-based methods to analyze the sensitivity of aerodynamic forces around bluff bodies at various Reynolds numbers \cite{meliga2014sensitivity}. Similarly, the adjoint-based optimization methods can be applied to obtain relaxation factor's sensitivity field. The deep RL algorithm may not be competitive when compared to adjoint methods which generally involve only two numerical simulations (one for the forward simulation, and one for the adjoint solution). Another shortcoming of deep RL algorithm is the \emph{a priori} specification of several hyperparameters (for example the update frequency of relaxation factor) that may affect the accuracy significantly. While there are some empirical remedies for this for supervised learning applications through hyperparameter search strategies, RL and its episode hungry nature make this step intractable in the absence of extremely large compute resources. However, it is important to recognize that a trained RL policy, while costlier in terms of off-line model evaluations, can be deployed across different on-line environments provided a representative set of simulations was run during its training. Therefore RL based simulation control has possibilities with regard to computational cost amortization over the course of an extended numerical simulation campaign.}

\textcolor{blue}{More generally, adjoint-based methods are very difficult to apply for the optimization of time-averaged quantities in turbulent flows (such as in LES applications). This is due to the chaotic nature of turbulence, and a small change in the initial condition can cause exponentially diverging adjoint solutions as soon as the length of the adjoint simulation exceeds the predictability time scale \cite{wang_gao_2013}. Therefore, the deep RL methods can be an alternative to adjoint-based methods in terms of their application.}

\section{Concluding Remarks}\label{sec:conclusion}
In this study, we have investigated the application of distributed deep RL to automatically update the computational hyperparameters of numerical simulations, which is a common task in many scientific applications. We illustrate our framework for two problems. The first problem is tasked with sustaining chaos in chaotic systems, and the second problem is the acceleration of convergence for steady-state CFD solver. In the first example, we demonstrate that the policy learned by an RL agent trained using proximal policy optimization (PPO) algorithm can automatically adjust the system's parameters to maintain chaos in transiently chaotic systems. The problem of controlling relaxation factors in steady-state CFD solvers is different from the standard control problem. In a standard control problem, the target value is known in advance and this target value is used to change action variables in such a way that the state of the system is guided towards the target state. However, the problem of accelerating convergence of steady-state CFD solvers is different in a way that the target point, i.e., the converged solution is not known. Therefore, RL is a suitable algorithm for this problem to discover the decision-making strategy that will dynamically update the relaxation factor with an objective to minimize the number of iterations required for convergence. Our results of numerical experiments indicate that the learned policy leads to acceleration in the convergence of steady-state CFD solver and the learned policy is generalizable to unseen boundary conditions and different geometries.      

Despite the relative simplicity in terms of the RL problem formulation and its application to backward-facing step example, this framework can be readily extended for more complex multiphysics systems. The state and the action space of an RL agent can be augmented to tackle complexities in these systems. \textcolor{blue}{Our future studies will revolve around efficient state construction that can optimally balance information about the environment and computational cost. This shall be crucial for large computational hyperparameter spaces in more challenging problems which will necessitate information from various flow field variables.} One of the major challenges in using RL for scientific simulations is the computational time due to a slow environment, and this challenge can be addressed by utilizing distributed environment evaluation along with high-performance computing to reduce the CPU time of the solver. \textcolor{blue}{A thorough study of asynchronous RL techniques for large simulations in distributed environments is also important for deploying our ideas to practical CFD applications.}

\section*{Acknowledgement}
We acknowledge the helpful discussions with Venkat Vishwanath and the Argonne Leadership Computing Facility Datascience team. Also, we thank Omer San for his insightful comments on this manuscript. This material is based upon work supported by the U.S. Department of Energy (DOE), Office of Science, Office of Advanced Scientific Computing Research, under Contract DE-AC02-06CH11357. This research was funded in part and used resources of the Argonne Leadership Computing Facility, which is a DOE Office of Science User Facility supported under Contract DE-AC02-06CH11357. This paper describes objective technical results and analysis. Any subjective views or opinions that might be expressed in the paper do not necessarily represent the views of the U.S. DOE or the United States Government.

\bibliographystyle{unsrt} 
\bibliography{ref}

\end{document}